\documentclass[%
reprint,
amsmath,amssymb,
aps,
prd,
nofootinbib, 
]{revtex4-2}

\usepackage{graphicx}
\usepackage{dcolumn}
\usepackage{bm}
\usepackage{xcolor}
\usepackage{lipsum}
\usepackage{hyperref}

\usepackage{soul}

\usepackage[normalem]{ulem}

\newcommand{\restr}[2]{{
		\left.\kern-\nulldelimiterspace
		#1 
		\vphantom{\big|}\,
		\right|_{#2}
	}}

\begin{document}

\title{Local clustering of relic neutrinos with kinetic field theory}

\author{Emil Brinch Holm}
\email{ebholm@phys.au.dk}
\affiliation{
	\vspace{0.2cm}
	Department of Physics and Astronomy, Aarhus University, DK-8000 Aarhus C, Denmark
}

\author{Isabel M. Oldengott}
\email[]{isabel.oldengott@uclouvain.be}
\affiliation{
	\vspace{0.2cm}
	Centre for Cosmology, Particle Physics and Phenomenology,
Université catholique de Louvain, Louvain-la-Neuve B-1348, Belgium
}

\author{Stefan Zentarra}
\email{szentarra@ethz.ch}
\affiliation{
	\vspace{0.2cm}
	Institute for Theoretical Physics, ETH Zurich, Wolfgang-Pauli-Str.~27, 8093 Zurich, Switzerland
}

\date{\today}

\begin{abstract}
	The density of relic neutrinos is expected to be enhanced due to clustering in our local neighbourhood at Earth. We introduce a novel analytical technique to calculate the neutrino overdensity, based on kinetic field theory. Kinetic field theory is a particle-based theory for cosmic structure formation and in this work we apply it for the first time to massive neutrinos. The gravitational interaction is expanded in a perturbation series and we take into account the first-order contribution to the local density of relic neutrinos. For neutrino masses that are consistent with cosmological neutrino mass bounds, our results are in excellent agreement with state-of-the-art calculations.
\end{abstract}

\maketitle

\section{Introduction}

While at large scales the formation of structures in our Universe can be described within the well-established framework of cosmological linear perturbation theory, on smaller scales our predictions almost entirely rely on $N$-body simulations~\cite{Euclid:2022qde}. In general, $N$-body simulations reproduce observed structures in remarkable detail and allow to derive testable features. 
Nonetheless, such simulations, unlike analytical approaches, are very computationally demanding and may lack physical interpretability.
The most desirable state would be an interplay between $N$-body simulations and (semi-)analytical methods. Unfortunately, analytical methods allow reliable predictions up to scales only slightly beyond the linear scale~\cite{Carlson:2009it}. An interesting and relatively recent approach to structure formation is provided by kinetic field theory (KFT)~\cite{Bartelmann2015traj,bartelmann2016KFT,bartelmann2019cosmic,Bartelmann:2020kcx,Konrad:2022fcn,Heisenberg:2022uhb,Pixius:2022hqs}. The main effort of KFT in cosmology has so far been the calculation of the matter power spectrum~\footnote{Beyond that, an application to baryonic accoustic oscillations can be found in \cite{Kostyuk:2023gqb} and velocity power spectra have been studied in~\cite{Littek_PhDthesis}.}, showing promising results. The formalism, however, might seem relatively complicated and---being based on a path integral formulation of classical particles---is very different from standard analytical methods. As such, this recent development seems to have obtained relatively little attention. 
The idea of this work is to take one step back and test the reliability of KFT on a considerably less complicated problem, namely the calculation of the \emph{local overdensity of relic neutrinos}. As our final results look promising for realistic neutrino masses, this work can be seen as introducing a novel technique to approach this problem. 

The relic or cosmic neutrino background (C$\nu$B) can be seen as the neutrino analogue of the cosmic microwave background (CMB). Its presence has so far been confirmed only indirectly, by measurements of the number of relativistic degrees of freedom~$N_{\text{eff}}$ through the CMB anisotropy spectrum (e.g.~\cite{Planck:2018vyg}) and through light element abundances from big bang nucleosynthesis (BBN) (e.g.~\cite{Escudero:2022okz,Pitrou:2018cgg}). A direct detection would however still be the ultimate dream of every neutrino cosmologist. Originating back to temperatures $T \sim 1$ MeV, it would provide the earliest direct probe of the Universe that we have. 
Due to the extremely low energies of the relic neutrinos today, a direct detection is unfortunately \emph{extremely} challenging -- if even feasible at all~\cite{KATRIN:2022kkv}. Nevertheless, the ambitious PTOLEMY project~\cite{Betts:2013uya} has taken on the challenge of a first direct detection of the C$\nu$B. A crucial parameter for PTOLEMY (and for any other future attempt to measure the C$\nu$B) is the local neutrino density which directly impacts the expected event rate (by its impact on the Tritium $\beta$-decay rate induced by neutrino capture). While the homogeneous and isotropic background neutrino density is expected to be around 56~cm$^{-3}$ per flavour (and as many for anti-neutrinos), the local neutrino density is in general expected to be enhanced because neutrinos start clustering when they become non-relativistic. The current strongest experimental constraint on the local neutrino overdensity from the KATRIN experiment is $< 1.1 \times 10^{11}$ (95\%~CL)~\cite{KATRIN:2022kkv}. A comparable or somewhat stronger bound is expected from future experiments for cosmogenic neutrinos \cite{Brdar:2022kpu}.

From the theoretical side, this problem has first been approached in~\cite{Singh:2002de} where it was pointed out that calculating the local neutrino clustering does not require full $N$-body simulations but can be simplified significantly by what we here dub the \emph{$N$-one-body approximation}. Within this approximation, the relic neutrinos are treated as test particles in an external gravitational potential governed by cold dark matter (CDM) and baryons. In other words, neutrino clustering neither impacts the clustering of CDM, baryons nor itself. The validity of the $N$-one-body approximation was explicitly confirmed by~\cite{Brandbyge:2010ge} and is the basis for our work and all methods introduced in this section. Based on this assumption, \cite{Singh:2002de} used the linearized Vlasov equation to solve for the local neutrino density. \cite{Ringwald:2004np} then showed that the linearized Vlasov equation is only justified for sufficiently small neutrino overdensities by performing $N$-one-body simulations. Instead of simultaneously following the evolution of $N$ particles (neutrinos and CDM) as in a conventional $N$-body simulation, the application of the $N$-one-body approximation allows the tracking of one particle at a time through $N$ independent simulations, which reduces the computation time significantly. The same method---with various improvements on the modelling of the gravitational potential---has been adopted in~\cite{deSalas:2017wtt,Zhang:2017ljh}. \cite{Ringwald:2004np,deSalas:2017wtt,Zhang:2017ljh} all assumed a spherically symmetric gravitational potential to keep the computational effort in a reasonable scale. Most recently, \cite{Mertsch:2019qjv} has extended the computation of the local overdensity to a non-spherical gravitational potential and applied the \emph{back-tracking technique}. Here, the trajectories of the neutrinos ending up \emph{here} and \emph{now} are tracked back by reversing the time direction in the equations of motion. The local overdensity is then  derived by assigning a statistical weight to each trajectory. Long story short, the final conclusion of~\cite{Mertsch:2019qjv} 
is that the local relic neutrino overdensity is $[0.5\%, 12\%, 50\%, 500\%]$ for neutrino masses $[10, 50, 100, 300] \, \text{meV}$. For neutrino masses not in tension with cosmological neutrino mass bounds~\cite{Planck:2018vyg} the overdensity is hence expected to be $< 50\%$.

While we thus already have a theoretical prediction for the local neutrino overdensity, the problem at hand offers the perfect conditions to test the reliability of KFT: Due the application of the $N$-one-body approximation, the KFT formalism reduces to a formalism of one-particle quantities. Therefore, we avoid two major complications arising in the context of calculations of the matter power spectrum with KFT, namely correlated initial conditions~\cite{bartelmann2016KFT,Bartelmann:2016kmx,Fabis:2017kju} and inter-particle (gravitational) interactions~\cite{bartelmann2016KFT,Heisenberg:2022uhb,Pixius:2022hqs}. Furthermore, as just summarized, the problem has been well explored with other methods which offers the opportunity to directly compare our results with state-of-the art methods.

\section{The formalism}

The formalism of KFT for an ensemble of $N$ particles has been developed thoroughly in~\cite{bartelmann2019cosmic} (plus references within) and~\cite{Bartelmann:2020kcx,Konrad:2022fcn,Heisenberg:2022uhb,Pixius:2022hqs}. It is intuitively clear (and can be shown rigorously) that the use of the $N$-one-body approximation justifies the formulation of the problem in terms of one-particle quantities. While the formalism presented in this section looks significantly simpler than in most of the literature on KFT, we note that all equations can be derived from the formalism as introduced in~\cite{Heisenberg:2022uhb}.

In a nutshell, KFT is a path integral formulation of classical mechanics. Its central object is the generating functional, averaged over initial conditions, 
\begin{align}
   & Z \left[ \Vec{J} \right] = \frac{1}{N (2 \pi)^{3}} \int \mathrm{d}^3 x^{(i)} \, \mathrm{d}^3 \, p^{(i)} \frac{1}{\exp \left(p^{(i)}/T_0 \right)+1} \nonumber \\
    & \hspace{1cm} \times \exp \left( i \int_{z_i}^{z_f} \mathrm{d} z' \,\Vec{J}(z') \cdot \Vec{x}^{\,s} \left(z';\Vec{x}^{(i)},\Vec{p}^{(i)} \right) \right)  \, . \label{eq:generating_functional} 
\end{align}
Here, the first line depicts the averaging over initial \emph{comoving} positions~$\vec{x}^{(i)}$ and initial conjugate momenta~$\vec{p}^{(i)}$. For the initial phase-space distribution we assume a relativistic, homogeneous and isotropic Fermi-Dirac distribution, with $T_0$ denoting the neutrino temperature today. We denote the norm of the momentum by $p^{(i)}$. $\vec{x}^{s}$ is the solution to the Hamiltonian equations of motion and $\vec{J}$ is the so-called source function\footnote{Note that more generally, one can also introduce a momentum source function as in~\cite{bartelmann2016KFT,Heisenberg:2022uhb}. However, in this work we are interested in the (momentum-independent) number density of neutrinos, so we can immediately set the momentum source function to zero in eq.~\eqref{eq:generating_functional}.}. Note that we explicitly choose redshift $z$ as our time coordinate. Physical observables can be derived from the generating functional by means of functional differentiation with respect to the source function. As such, the number density in Fourier space can be derived according to
\begin{equation}
    n(\vec{k},z) = N  \left.\exp\left( - i \vec{k} \cdot \frac{\delta}{i \delta \vec{J}(z)} \right) \, Z\left[ \vec{J} \right] \right\vert_{\vec{J}= 0}\, . 
    \label{eq:number_Fourier}
\end{equation}
In order to evaluate eq.~\eqref{eq:number_Fourier} let us derive the equations of motion in the following. The Lagrangian with redshift $z$ as a time coordinate is given by
\begin{align}
	\mathcal{L}\left(\vec{x},\frac{\mathrm{d} \vec{x}}{\mathrm{d} z},z \right) =& -\frac{m_\nu}{2} \frac{H(z)}{(1+z)} \left(\frac{\mathrm{d} \vec{x}}{\mathrm{d} z}\right)^2 + \frac{m_\nu \varphi\left(\vec{x},z\right)}{(1+z) H(z)} \,.
 \label{eq:Lagrangian} 
\end{align}
We assume a two-component Universe for the Hubble expansion rate $H(z)=H_0 \sqrt{ \Omega_{m,0} (1 + z)^3 + \Omega_\Lambda}$, which is justified as neutrino clustering happens at sufficiently late times. Here $\Omega_{m,0}$ and $\Omega_{\Lambda}$ are the matter density and cosmological constant density parameters today, and $H_0$ is the Hubble rate today.
The effective gravitational potential $\varphi$ on the RHS of eq.~\eqref{eq:Lagrangian} fulfills the comoving Poisson equation (see e.g.\ the appendix of~\cite{Bartelmann2015traj}), 
\begin{align}
	\Delta \varphi(\vec{x}, z) = \frac{4\pi G}{(1+z)^2} \rho(\vec{x},z) \, ,
 \label{eq:Poisson}
\end{align}
where $\rho$ is the matter density perturbation.

The conjugate momentum to the comoving position is
\begin{align}
	\vec{p} = \frac{\partial \mathcal{L}}{\partial \left(\frac{\mathrm{d} \vec{x}}{\mathrm{d} z}\right)} = - m_\nu \frac{H(z)}{(1+z)} \frac{\mathrm{d} \vec{x}}{\mathrm{d} z} \,,
\end{align}
which leads to the Hamiltonian
\begin{align}
	\mathcal{H}\left(\vec{x}, \vec{p}, z\right) 
	= - \frac{1+z}{2 m_\nu H(z)} \vec{p}^{\,2} - \frac{m_\nu \varphi \left(\vec{x},z\right)}{(1+z) H(z)} \,
\end{align}
and hence the equations of motion
\begin{align}
	\frac{\mathrm{d} \vec{x}}{\mathrm{d} z} &= \frac{\partial \mathcal{H}}{\partial \vec{p}} = - \frac{1+z}{m_\nu H(z)} \vec{p} \, , \label{eq:eoms_1} \\
	\frac{\mathrm{d} \vec{p}}{\mathrm{d} z} &= -\frac{\partial \mathcal{H}}{\partial \vec{x}} = \frac{m_\nu }{(1+z) H(z)} \frac{\partial \varphi \left(\vec{x},z\right)}{\partial \vec{x}} \, .
	\label{eq:eoms_2}
\end{align}
If the equations of motion~\eqref{eq:eoms_1}-\eqref{eq:eoms_2} had an analytical solution, we could now derive an exact analytical expression for the number density~\eqref{eq:number_Fourier} in terms of an integral over initial phase-space. As this is not the case, we follow the perturbative treatment of the generating functional outlined in~\cite{Heisenberg:2022uhb}. For that purpose, we write down the formal solution of eqs.~\eqref{eq:eoms_1}-\eqref{eq:eoms_2},
\begin{equation}
\begin{aligned}
    \vec{x}^{\,s}(z) =& \vec{x}^{(i)} + g(z,z_i) \vec{p}^{(i)} \\
    & + m_{\nu} \int_{z_i}^z \mathrm{d}z' \, \frac{g(z,z')}{(1+z') H(z')} \frac{\partial \varphi(\vec{x}^{\,s}(z'))}{\partial \vec{x}} , \\
    \vec{p}^{\,s}(z) =& \vec{p}^{(i)} + m_{\nu} \int_{z_i}^z \mathrm{d}z' \, \frac{1}{(1+z') H(z')} \frac{\partial \varphi(\vec{x}^{\,s}(z'))}{\partial \vec{x}} \, 
    \label{eq:formal_solution}
\end{aligned}
\end{equation}
with the Green's function 
\begin{align} \label{eq:defn_propagator}
    g(z_2,z_1) &= -\int_{z_1}^{z_2} \mathrm{d} z' \, \frac{1+z'}{m_\nu H(z')} \, \\
    & \hspace{-1.5cm} = \left[ \frac{ (1 + z )^2}{2 H_0 m_\nu  \sqrt{\Omega_\Lambda}} \, \, {}_2F_1\left(\frac{1}{2},\frac{2}{3};\frac{5}{3};-\frac{\Omega_{m,0}}{\Omega_\Lambda} (z+1)^3\right) \right]_{z_2}^{z_1} \,, 
    \label{eq:free_propagator}
\end{align}
where ${}_2F_1$ refers to the Gaussian hypergeometric function.
Following the derivation in~\cite{Heisenberg:2022uhb} (for the case of an $N$-particle ensemble), one can then write the generating functional in eq.~\eqref{eq:generating_functional} as
\begin{align}
 Z \left[\vec{J} \right] &= \frac{1}{N (2 \pi)^{3}} \! \int \mathrm{d}^3 x^{(i)} \mathrm{d}^3 p^{(i)} \frac{1}{\exp \left(p^{(i)}/T_0 \right)+1} \label{eq:phase-space-integration} \\ 
& \hspace{-0.8cm} \times \exp \! \left( \! i \! \!\int_{z_i}^{z_f} \! \! \! \! \mathrm{d} z'' \!\! \int_{z_i}^{ z''} \! \! \! \! \mathrm{d} z' \frac{m_{\nu} \, g( \!z''\!,\!z'\!)}{(\! 1\! + \!z'\!) H( \! z' \!)} \vec{J}(\! z'' \! ) \! \cdot \! \frac{\partial \varphi}{\partial \vec{x}} \! \left( \! \frac{\delta}{i\delta \vec{J}(\! z' \!)}\!, \!z' \! \right) \! \! \right) \label{eq:interaction_operator} \\*
& \times \exp \left( i \int_{z_i}^{z_f} \! \mathrm{d} z' \, \vec{J}(z') \! \cdot \! \left( \vec{x}^{(i)} \! + \! g(z',z_i) \vec{p}^{(i)} \right) \right) \label{eq:Z0} \\
& \equiv   \exp \left( i \hat{S}_I \right) Z_0\left[\vec{J} \right] \, .
\end{align}
Here, we identified eq.~\eqref{eq:Z0} (including the integration over initial conditions in eq.~\eqref{eq:phase-space-integration}) as the \emph{free generating functional} $Z_0$ and eq.~\eqref{eq:interaction_operator} as an interaction operator~$\exp (i \hat{S}_I)$  acting on it. Applying the series definition of the exponential~$\exp (i \hat{S}_I)$ leads to a perturbation series for the generating functional~\eqref{eq:generating_functional} and in turn to a perturbation series for the number density,
\begin{equation}
    n(\vec{x},z) = \bar{n}(\vec{x},z) + \delta n^{(1)}(\vec{x},z) + \ldots \, .
    \label{eq:perturbed_number}
\end{equation}
In this work, we restrict our analysis to KFT in first-order perturbation theory. 
As discussed in sec.~4.4 of~\cite{Heisenberg:2022uhb}, the particle trajectories from the first-order generating functional correspond to trajectories within the \emph{Born approximation}. To be more explicit, those are trajectories in eq.~\eqref{eq:formal_solution} where the gradient of the gravitational potential is evaluated along the free trajectory.

At zeroth order, applying the functional derivative in eq.~\eqref{eq:number_Fourier} on the free generating functional directly gives the standard expression for the homogeneous and isotropic background number density, i.e. 
\begin{align}
    \bar{n}(z) =  \frac{1}{(2\pi)^3} \int \mathrm{d}^3 p^{(i)} \frac{1}{\exp \left( p^{(i)}/T_0 \right) + 1} \, .
\end{align}
The derivation in appendix \ref{sec:appendix} shows that the first order contribution to the local relic neutrino number density is 
\begin{align}
& \delta n^{(1)}\left( \vec{x}, z \right) 
= - \frac{4\pi G }{(2 \pi)^3} \int \mathrm{d}^3 p^{(i)}  \frac{1}{\exp \left( p^{(i)}/T_0 \right) +1 } \nonumber \\ 
&      \times \! \int_{z_i}^{z} \! \mathrm{d} z' \! \frac{m_\nu \, g(z,z')}{(1+z')^3 H(z')} \, \rho \! \left( \! \vec{x} \! - \! g(z,z') \vec{p}^{(i)} ,z'\right) \, .
\label{eq:number_first_order}
\end{align}
The above expression has a transparent interpretation: At first-order, the contribution to the relic neutrino overdensity is obtained by integrating the external matter density along the free trajectory back in time from the position $\vec{x}$.

\section{Results}
Eq.~\eqref{eq:number_first_order} and the numerical calculation thereof form the main result of this work. We compare it to the most state-of-the art technique to calculate the local neutrino overdensity, i.e., the back-tracking method applied by~\cite{Mertsch:2019qjv}. While~\cite{Mertsch:2019qjv} showed that in general non-spherical contributions to the gravitational potential (in particular the Virgo cluster and baryonic components of the Milky way) have a significant impact on the local neutrino overdensity, 
the aim of this work is to provide a test of KFT and a comparison between the two methods. For this purpose, the modelling of the non-spherical potential as described in~\cite{Mertsch:2019qjv} would present an unnecessary complication\footnote{Note that spherical symmetry has also been assumed in all works previous to~\cite{Mertsch:2019qjv}, i.e.~\cite{Ringwald:2004np,deSalas:2017wtt,Zhang:2017ljh}.}. We therefore restrict our work to the case of a spherically symmetric NFW potential \cite{Navarro:1995iw} representing the dark matter halo of the Milky Way and model it \emph{exactly} as described in~\cite{Mertsch:2019qjv}. We truncate the NFW profile at its virial radius $R_\mathrm{vir} (z)$, 
\begin{align}
    \rho \! \left( \vec{x} ,z\right) =
        \begin{cases}
            \frac{\rho_0(z) R_s(z)^3}{r(r + R_s(z))^2}, &r<R_\mathrm{vir}(z) \,,\\
            0, &r>R_\mathrm{vir}(z) \,,
        \end{cases}
        \label{eq:NFW-profile}
\end{align}
where $\rho_0 (z)$ denotes the central density and $R_s(z)$ the scale radius. We refer the reader to~\cite{Mertsch:2019qjv} for the time evolution of $\rho_0(z)$, $R_s(z)$ and $R_{\text{vir}}(z)$ as well as all numerical values that are needed to fully characterize the density profile.
As in~\cite{Mertsch:2019qjv}, we assume the Earth to be located at a distance of $8.2\,\text{kpc}$ from the galactic center~\cite{doi:10.1146/annurev-astro-081915-023441} and assume the cosmological parameters $\Omega_{m,0}=0.3111$, $\Omega_\Lambda = 0.6889$ and $H_0=67.77 \, \text{km s}^{-1}\text{Mpc}$.

Numerically, we evaluate the integral~\eqref{eq:number_first_order} as a four-dimensional iterated integral. However, by choosing a coordinate system in which $\vec{x}=x  \vec{e}_z$, one can show that the magnitude of the vector $\vec{x} - g(z,z'') \vec{p}^{(i)}$ is independent of the azimuthal angle of the initial momentum $\vec{p}^{ (i)}$, reducing the dimensionality of the integral by one. We take the redshift integral as the outermost integral and evaluate the hypergeometric function, occurring in the Green's function~\eqref{eq:free_propagator}, with a $(12,12)$ minimax rational approximation~\cite{numerical_recipes}. The redshift and polar integrals are evaluated using adaptive quadrature with a 15 point Gauss-Kronrod formula~\cite{KahanerDavid1989Nmas} and the $\mathrm{d} p^{(i)}$ integral is performed using Gauss-Laguerre quadrature~\cite{numerical_recipes} since it is weighted by the initial Fermi-Dirac distribution. Without parallellization, the integral only takes on the order of $10$ ms to evaluate\footnote{Our code is publicly available at \texttt{https://github.com/EBHolm/KFT-Neutrinos} on the branch with the arXiv ID of this paper.}. This is \emph{many} orders of magnitude faster than the back-tracking method, which takes about 120 minutes, according to~\cite{Mertsch:2019qjv}. However, our assumption of spherical symmetry reduces the dimensionality of the integral~\eqref{eq:number_first_order} compared to the computation of~\cite{Mertsch:2019qjv}, which of course makes a one-to-one comparison of the computation times difficult.  

\begin{figure}[t]
    \includegraphics[width=\columnwidth]{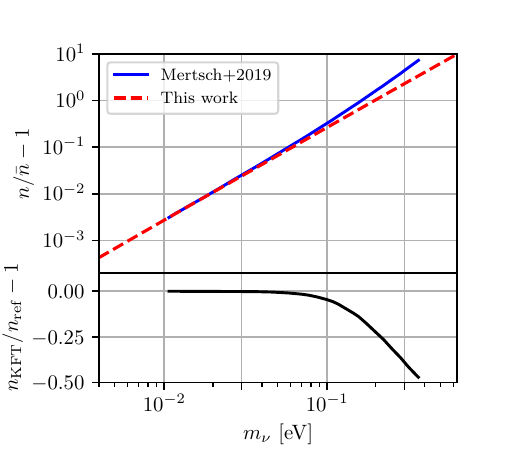}
    \caption{\label{fig:mass_first} Local relic neutrino overdensity from first-order KFT according to eq.~\eqref{eq:number_first_order} (red dashed) as a function of neutrino mass $m_{\nu}$, in comparison to the results of~\cite{Mertsch:2019qjv} (blue solid). The lower panel shows the relative difference between both methods.}
\end{figure}

Figure~\ref{fig:mass_first} shows the estimates of the local relic neutrino overdensity $n/\bar{n} - 1$ from eq.~\eqref{eq:perturbed_number} and eq.~\eqref{eq:number_first_order} compared to the state-of-the-art results of~\cite{Mertsch:2019qjv} as a function of the neutrino mass $m_\nu$ (the blue line corresponds to the line dubbed as ``NFW'' in fig.~3 of~\cite{Mertsch:2019qjv}). The lower panel shows the relative difference between the results of both methods. The two computations are seen to agree at the percent level for masses up to $\approx 0.1$ eV, while the first-order KFT result underestimates the neutrino clustering at larger masses. At $m_{\nu}=0.3$ eV the difference between both methods is around $47 \%$. 

The fact that the result from first-order KFT \emph{underestimates} the overdensity can be well understood by keeping in mind that it corresponds to Born trajectories that underestimate the deviation from straight lines (or free trajectories). This can also be seen as an underestimation of the gravitational force and it therefore comes as no surprise that the neutrino clustering is underestimated as well. For low neutrino masses this effect is less severe because the true trajectories are less curved (due to the higher velocities) and the Born approximation performs better. Furthermore, for high neutrino masses one can in general expect neutrinos to follow bound trajectories, which KFT is unable to reproduce at first order. 

\section{Conclusion}

In this work, we have for the first time applied KFT to calculate the local overdensity of relic neutrinos.
For neutrino masses consistent with cosmological mass bounds, first-order KFT shows excellent agreement with the state-of-the art calculation by~\cite{Mertsch:2019qjv} that applies the back-tracking method. For larger neutrino masses ($m_\nu \gtrsim 0.1$ eV) the overdensity is underestimated (up to around $47\%$ at $m_{\nu}=0.3$~eV). The method introduced in this work is a priori not restricted to relic neutrinos but can be applied to other light particles, under the assumption that the $N$-one-body approximation holds.

We focused on a spherically symmetric NFW potential governed by dark matter. This is justified as our work should be rather seen as a proof-of-principle for the reliability of KFT---and not so much as a precision calculation of the local clustering of relic neutrinos which would require accounting for baryonic contributions from the Milky Way and the Virgo cluster (see~\cite{Mertsch:2019qjv}). 
Though the calculation of the local relic neutrino density is not a cosmological problem limited by computation time, it is remarkable that the numerical evaluation of eq.~\eqref{eq:number_first_order} only takes about 10 ms (per neutrino mass). To put this into context, the back-tracking method takes around 120 minutes according to~\cite{Mertsch:2019qjv}. However, caution has to be allowed with this direct comparison of computation times as the generalization to a non-symmetric gravitational potential like in~\cite{Mertsch:2019qjv} could increase the computation time in a non-trivial way. However, even in comparison to previous works based on $N$-one-body simulations~\cite{Ringwald:2004np,deSalas:2017wtt,Zhang:2017ljh}, which also assumed spherical symmetry, the first-order KFT computation is considerably faster. On the other hand, it was pointed out early on in~\cite{Ringwald:2004np} that for relatively low overdensities (and hence for what we today consider as realistic neutrino masses) solving the linearized Vlasov equation already provides reliable results. Indeed, the contribution from first-order KFT in eq.~\eqref{eq:number_first_order} has a form similar to the \emph{linear response} expression in~\cite{Ringwald:2004np} (or \cite{Chen:2020kxi} in the context of $N$-body simulations including massive neutrinos), where neutrinos respond linearly to an external gravitational potential. Note however that in case of the linearized Vlasov equation the perturbation parameter is the phase-space density while in KFT it is the gravitational interaction. It is a priori not clear how both of these approaches compare. In a future work we will compare KFT to the linearized Vlasov equation (and its numerical performance). We believe that such a comparison is also relevant for a general comparison between KFT and standard cosmological perturbation theory~\cite{Kozlikin_2021}. Furthermore, we leave the investigation of higher-order contributions to the clustering for future work.

In conclusion, KFT reproduces the standard results of a homogeneous and isotropic background neutrino density and performs remarkably well at first order. For the specific calculation of the local clustering of relic neutrinos and for neutrino masses consistent with cosmological mass bounds, the method introduced in this work is competitive with the state-of-the art technique based on back-tracking~\cite{Mertsch:2019qjv}.


\acknowledgments
We thank M.~Bartelmann, S.~Hannestad, L.~Heisenberg, P.~Mertsch, T.~Tram and Y.\,Y.\,Y.~Wong for interesting and fruitful discussions.
I.\,M.\,O.\ acknowledges support by Fonds de la recherche scientifique (FRS-FNRS). E.\,B.\,H.\ was supported by a research grant (29337) from VILLUM FONDEN.

\begin{widetext}
\appendix

\section{Derivation of first-order contribution to the number density}
\label{sec:appendix}

In this appendix we derive eq.~\eqref{eq:number_first_order} for the first-order contribution to the local relic neutrino number density from KFT. For reasons of compactness, in the following we abbreviate the integral over initial conditions as
\begin{equation}
    \int \mathrm{d} \Gamma \dots = \frac{1}{N (2 \pi)^{3}} \! \int \mathrm{d}^3 x^{(i)} \mathrm{d}^3 p^{(i)} \frac{1}{\exp \left(p^{(i)}/T_0 \right)+1} \dots \,.
\end{equation}
The first-order generating functional can be written as 

    \begin{align}
        Z^{(1)} [\vec{J}] &= i  \! \int \! \mathrm{d} \Gamma   \! \int_{z_i}^{z_f} \! \! \mathrm{d} z'' \! \int_{z_i}^{z''} \! \! \mathrm{d} z' \! \frac{m_\nu \, g(z'',z')}{(1 \!+\! z') H(z')} \, \vec{J}(z'') \! \cdot \! \frac{\partial \varphi}{\partial \vec{x}} \left(\! \frac{\delta}{i\,\delta \vec{J}(z')},z' \!\right) \exp \left( \! i \int_{z_i}^{z_f} \! \mathrm{d} z' \vec{J}(z') \! \cdot \! \left( \vec{x}^{(i)} \! + \! g(\!z'\!,\!z_i\!) \vec{p}^{(i)} \right) \! \right) \\ 
        &= i  \! \int \! \mathrm{d} \Gamma  \! \int_{z_i}^{z_f} \! \mathrm{d} z'' \! \! \int_{z_i}^{z''} \! \! \mathrm{d} z' \! \frac{m_\nu  g(\!z''\!,\!z'\!)}{(1 \!+ \!z') H(z')} \vec{J}(z'') \! \cdot \! \frac{\partial \varphi}{\partial \vec{x}} \left(\! \vec{x}^{(i)} \! + \! g(\!z'\!,\!z_i\!) \vec{p}^{(i)} ,z' \!\right) \exp \! \left( \! i \! \int_{z_i}^{z_f} \! \mathrm{d} z' \vec{J}(z') \! \cdot \! \left( \! \vec{x}^{(i)} \! + \! g(\!z'\!,\!z_i\!) \vec{p}^{(i)} \right) \! \right) ,
    \end{align}
    
where we have applied eq.~(11) of~\cite{Heisenberg:2022uhb}. This can now be inserted into eq.\eqref{eq:number_Fourier} for the number density in Fourier space,

\begingroup
\allowdisplaybreaks
    \begin{align}
         \delta n^{(1)}\left( \vec{k}, z \right) 
=& \int \mathrm{d}\Gamma  \exp\left( - i \vec{k} \cdot \frac{\delta}{i \delta \vec{J}(z)} \right)   \left( i \int_{z_i}^{z_f} \mathrm{d} z'' \int_{z_i}^{z''} \mathrm{d} z' \, \frac{m_\nu \, g(z'',z')}{(1+z') H(z')} \, \vec{J}(z'') \cdot \frac{\partial \varphi}{\partial \vec{x}} \left(\vec{x}^{(i)} \! + \! g(z',z_i) \vec{p}^{(i)} ,z' \right) \right) \nonumber\\*
	& \times \restr{\exp \left( i \int_{z_i}^{z_f} \mathrm{d} z' \, \vec{J}(z') \cdot \left( \vec{x}^{(i)} + g(z',z_i) \vec{p}^{(i)} \right) \right)}{\vec{J} = 0} \\
     =& \int \mathrm{d}\Gamma \int_{z_i}^{z_f} \mathrm{d} z'' \int_{z_i}^{z''} \mathrm{d} z' \, \frac{m_\nu \, g(z'',z')}{(1+z') H(z')} \, \left( i\vec{J}(z'') -i \vec{k} \delta_D(z-z'') \right) \cdot \frac{\partial \varphi}{\partial \vec{x}} \left(\vec{x}^{(i)} \! + \! g(z',z_i) \vec{p}^{(i)} ,z' \right) \nonumber\\*
	& \times \restr{ \exp\left( - i \vec{k} \cdot \frac{\delta}{i \delta \vec{J}(z)}\right) \exp \left( i \int_{z_i}^{z_f} \mathrm{d} z' \, \vec{J}(z') \cdot \left( \vec{x}^{(i)} + g(z',z_i) \vec{p}^{(i)} \right) \right)}{\vec{J} = 0} \label{eq:appendix_productRule}\\
 =& \int \mathrm{d}\Gamma \exp\left( - i \vec{k}  \cdot \left( \vec{x}^{(i)} + g(z,z_i) \vec{p}^{(i)} \right)  \right)   \left(  \int_{z_i}^{z} \mathrm{d} z' \, \frac{m_\nu \, g(z,z')}{(1+z') H(z')} \, (-i \vec{k}) \cdot \frac{\partial \varphi}{\partial \vec{x}} \left(\vec{x}^{(i)} \! + \! g(z',z_i) \vec{p}^{(i)} ,z' \right) \right) \, .
\end{align}
\endgroup
In~\eqref{eq:appendix_productRule} we applied a product rule giving rise to two terms of which the first vanishes upon setting $\vec{J} = 0$. The local relic neutrino density is derived upon Fourier transformation of this expression,
\begin{align}
 \delta n^{(1)}\left( \vec{x}, z \right) 
	    &= - \frac{1}{(2 \pi)^3} \int \mathrm{d}^3 p^{(i)}  \frac{1}{\exp \left( p^{(i)}/T \right) +1 }  \int \mathrm{d}^3 x^{(i)} \int \frac{\mathrm{d}^3 k}{(2\pi)^3} \exp\left( - i \vec{k} \cdot \left( \vec{x}^{(i)} + g(z,z_i) \vec{p}^{(i)} - \Vec{x} \right) \right)  \nonumber\\*
        &\qquad \times \int_{z_i}^{z} \mathrm{d} z' \, \frac{m_\nu \, g(z,z')}{(1+z') H(z')} \, i \vec{k} \cdot \frac{\partial \varphi}{\partial \vec{x}} \left( \vec{x}^{(i)} + g(z',z_i) \vec{p}^{(i)},z'\right) \\
	    &= - \frac{1}{(2 \pi)^3} \int \mathrm{d}^3 p^{(i)}  \frac{1}{\exp \left( p^{(i)}/T \right) +1 }  \int \mathrm{d}^3 x^{(i)} \, \delta_D\left( \vec{x} - g(z,z_i) \vec{p}^{(i)} - \vec{x}^{(i)} \right) \nonumber\\*
        &\qquad \times \int_{z_i}^{z} \mathrm{d} z' \, \frac{m_\nu \, g(z,z')}{(1+z') H(z')} \, \frac{\partial}{\partial \vec{x}}  \cdot \frac{\partial \varphi}{\partial \vec{x}} \left( \vec{x}^{(i)} + g(z',z_i) \vec{p}^{(i)} ,z'\right) \\
&	    = - \frac{1}{(2 \pi)^3} \int \mathrm{d}^3 p^{(i)}  \frac{1}{\exp \left( p^{(i)}/T_0 \right) +1 } \! \int_{z_i}^{z} \! \mathrm{d} z' \! \frac{m_\nu \, g(z,z')}{(1+z') H(z')} \, \frac{\partial^2 \varphi}{\partial \vec{x}^2} \! \left( \! \vec{x} \! - \! g(z,z') \vec{p}^{(i)} ,z'\right) \, .
\label{eq:number_first_order_appendix} 
\end{align}
Here, we applied the Fourier representation of the Dirac delta function and used the relation $g(z',z_i)-g(z,z_i)=g(z',z)=-g(z,z')$ from eq.~\eqref{eq:defn_propagator}. Inserting the comoving Poisson equation~\eqref{eq:Poisson} into eq.~\eqref{eq:number_first_order_appendix} directly gives eq.~\eqref{eq:number_first_order}.

\end{widetext}

\bibliography{literature}

\end{document}